\documentclass{article}
\usepackage{spconf,amsmath,graphicx}
\usepackage{amsfonts,bm}
\usepackage{url}

\usepackage{multicol}
\usepackage{multirow}
\usepackage{booktabs}
\usepackage{color}
\usepackage{xcolor}
\usepackage[normalem]{ulem}

\title{TEMPORAL KNOWLEDGE DISTILLATION FOR \\ ON-DEVICE AUDIO CLASSIFICATION}
\name{
    Kwanghee Choi\textsuperscript{$\dagger$}\thanks{\textsuperscript{$\dagger$}Equal contribution.}, Martin Kersner\textsuperscript{$\dagger$}, Jacob Morton\textsuperscript{$\dagger$}, Buru Chang\textsuperscript{*}\thanks{\textsuperscript{*}Corresponding author.}
}
\address{
    Hyperconnect, South Korea
}
\begin{document}
%\ninept
%
\maketitle
\begin{abstract}\label{sec:0_abstract}
Improving the performance of on-device audio classification models remains a challenge given the computational limits of the mobile environment.
Many studies leverage knowledge distillation to boost predictive performance by transferring the knowledge from large models to on-device models.
However, most lack a mechanism to distill the essence of the temporal information, which is crucial to audio classification tasks, or similar architecture is often required.
In this paper, we propose a new knowledge distillation method designed to incorporate the temporal knowledge embedded in attention weights of large transformer-based models into on-device models.
Our distillation method is applicable to various types of architectures, including the non-attention-based architectures such as CNNs or RNNs, while retaining the original network architecture during inference.
Through extensive experiments on both an audio event detection dataset and a noisy keyword spotting dataset, we show that our proposed method improves the predictive performance across diverse on-device architectures.
\end{abstract}


\section{Introduction}\label{sec:1_introduction}
With the ubiquity of real time communication, on-device audio understanding has received great attention.
On-device models have achieved comparable performance to much larger models on several tasks such as keyword spotting (KWS)~\cite{speechcommandsv2, Rybakov_2020}.
Nevertheless, compared to large models, on-device models still struggle with more complex tasks (e.g., audio event detection (AED)~\cite{fonseca2020fsd50k}).
Improving the performance of on-device models is challenging due to the restricted memory and computing resources in the mobile environment.

Several studies~\cite{FutamiIUMSK20,Lu2017KnowledgeDF} utilize knowledge distillation (KD)~\cite{hinton2015distilling} to tackle the problem described above, applying the knowledge of large models (teacher) to on-device models (student) without incurring any computational overhead at inference time.
Many on-device models commonly focus on the knowledge embedded in logits produced by the classification layer~\cite{FutamiIUMSK20,Lu2017KnowledgeDF,berg21_interspeech}, mainly because it can be easily applied even when the teacher and the student have dissimilar architectures.
However, temporal information, which is known to be beneficial in audio tasks~\cite{MunimIS19}, cannot be easily distilled when it is compressed into classifier logits~\cite{codert21}.

With the success of the transformer~\cite{vaswani2017attention}, recent studies~\cite{Zagoruyko2017pay,Chang2020IntraUtteranceSP} have focused on distilling the knowledge from self-attention maps, preserving the temporal information.
However, their methods are limited to transferring the knowledge between the same transformer-based architectures only, where even the smallest transformer variants remain computationally expensive for many mobile devices.
Also, it is not straightforward to transfer the knowledge of self-attention maps from the large transformer-based model to other architectures such as convolutional neural networks (CNNs) or recurrent neural networks (RNNs).

In this paper, we introduce a simple yet effective method that can distill the temporal knowledge from attention weights of large transformer-based models to on-device models of various architectures.
We first employ XLSR-wav2vec 2.0~\cite{conneau2020unsupervised} as a teacher model and extract attention weights from its self-attention maps.
We design the attention distillation loss for the on-device (student) models by attaching a simple attention layer only at training time to align the teacher and the student attention weights.
To evaluate the effectiveness of our proposed method, we conduct experiments on a real-world AED dataset (FSD50K~\cite{fonseca2020fsd50k}) and a noisy KWS dataset.
The noisy KWS dataset is constructed by injecting the existing KWS dataset samples (Google Speech Commands v2~\cite{speechcommandsv2}) into different speech-like noise audios~\cite{audioset17}, making the temporal information more important for classifying each specific target keyword.
Experimental results demonstrate that applying our method improves the predictive performance of various on-device models without any architectural changes on inference by distilling temporal information during training.

\section{Proposed Method}\label{sec:2_proposed_method}
\begin{figure*}[t] %%% t: top, b: bottom, h: here
\centering
\includegraphics[width=0.95\textwidth]{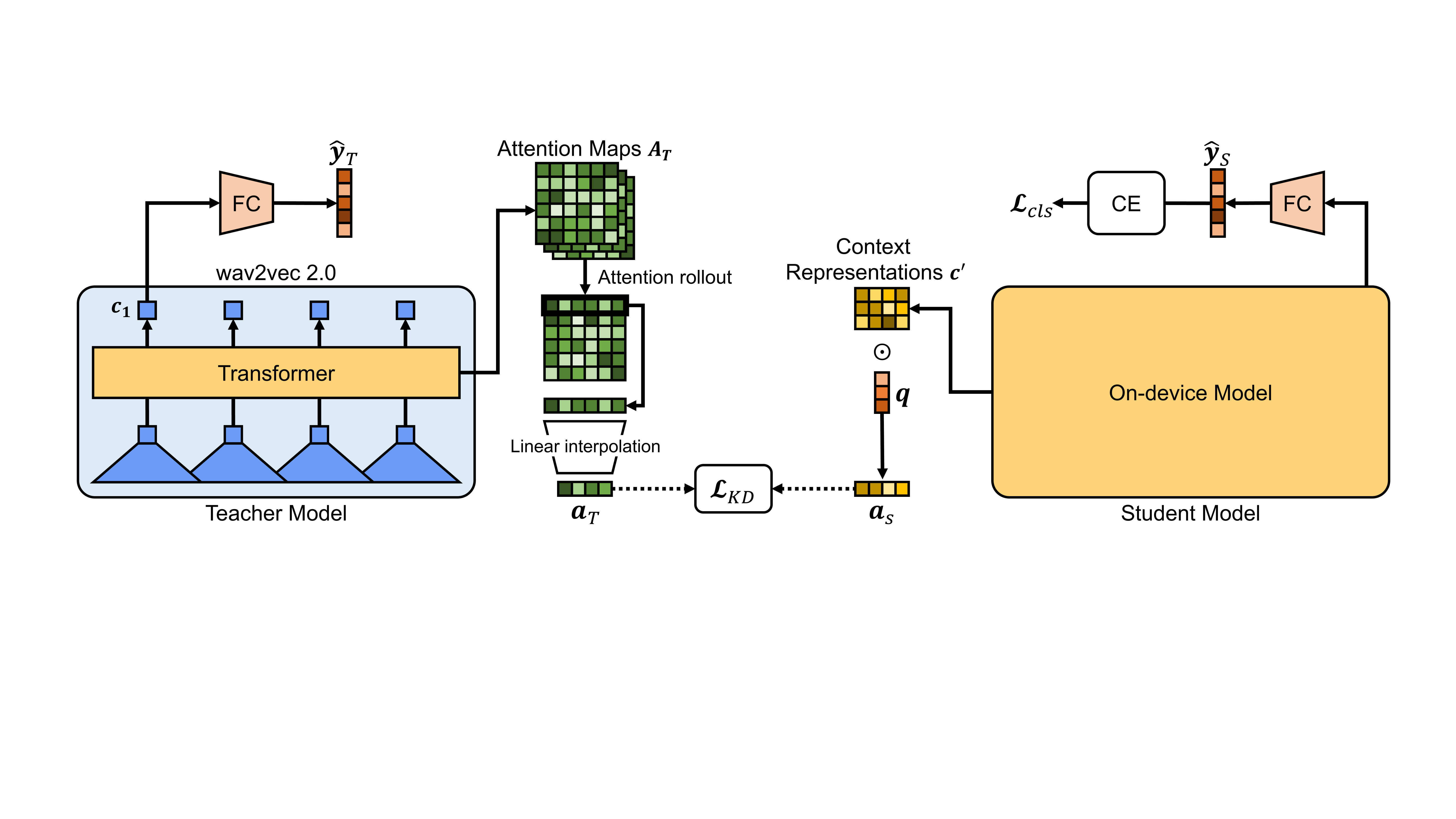}
\caption{
Illustration of our proposed method.
}
% \vspace{-0.5em}
\label{fig:1_model_architecture}
\end{figure*}
In this section, we first describe a large-scale transformer-based model that is used as a teacher model ($\S$\ref{subsec:2_1_teacher_model}) and on-device models employed as student models ($\S$\ref{subsec:2_2_student_model}).
We then introduce our method that transfers the temporal knowledge from self-attention of the teacher model to student models without any architectural changes during inference ($\S$\ref{subsec:3_zero_overhead_attention_distillation}).
Our proposed method is illustrated in Figure~\ref{fig:1_model_architecture}.

\subsection{Teacher: Large Transformer-based Model}\label{subsec:2_1_teacher_model}
We employ the XLSR-wav2vec 2.0~\cite{conneau2020unsupervised} as our teacher model $\mathcal{F}_T$, which is a large-scale transformer-based ASR model with state-of-the-art performance on multilingual ASR.
The teacher model translates a raw audio $x$ to latent representations for $n$ time-steps using a convolutional feature encoder.
The latent representations are passed through $m$ consecutive transformer layers to output context representations $\mathbf{c}_{i=1,\cdots,n}$, where $\mathbf{c}_i\in\mathbb{R}^{d}$ is the $d$-dimensional vector.

To perform audio classification, we attach a fully-connected layer to the output $\mathbf{c}$ of $\mathcal{F}_T$.
Similar to the fine-tuning of language models~\cite{devlin2019bert}, we feed only the first output $\mathbf{c}_1$ to the fully-connected layer.
$\mathcal{F}_T$ and the fully-connected layer are trained end to end on audio classification datasets (Details described in Sec. \ref{sec:3_experiment}).

\subsection{Student: Lightweight On-device Models}\label{subsec:2_2_student_model}
In this paper, we consider the following on-device audio classification models adopted by \cite{Rybakov_2020} as our student models $\mathcal{F}_S$: a simple RNN-based model (\textbf{LSTM-P})~\cite{DBLP:journals/jmlr/GersSS02}, a CNN-based model (\textbf{TC-ResNet})~\cite{Choi2019tcresnet}, a model that uses both CNN and RNN (\textbf{CRNN})~\cite{DBLP:conf/interspeech/ArikKCHGFPC17}, a model including an attention mechanism (\textbf{Att-RNN})~\cite{Andrade2018aneural}, and a multi-head variant of Att-RNN (\textbf{MHAtt-RNN})~\cite{Rybakov_2020}.
For the student models, we pass the raw audio to the MFCC-based feature encoder.
The student models extract the context representations $\mathbf{c'}_{i=1,\cdots,n'}$ for $n'$ time steps, where $\mathbf{c'}_i\in\mathbb{R}^{d'}$ is the $d'$-dimensional vector.
Note that the student models have different sizes $n'$ and $d'$ depending on their architecture.

We integrate the attention mechanism that extracts attention weights $\mathbf{a}_{S}\in\mathbb{R}^{n'}$ from every student model except Att-RNN and MHAtt-RNN architectures which already include the attention mechanism.
The extracted student attention weights act as recipients for transferring high-level knowledge from the teacher model $\mathcal{F}_T$.
The attention weights are computed by applying the softmax function on the inner-product of the context representations $\mathbf{c'}$ and a query $\mathbf{q}\in\mathbb{R}^{d'}$ as follows: $\mathbf{a}_S=\text{softmax}(\mathbf{c'}_i\cdot\mathbf{q})_{i=1}^{n'}$.
There are many strategies for designing the query $\mathbf{q}$, e.g., random initialization~\cite{yang2016hierarchical} or projection of the medium context representation~\cite{Andrade2018aneural}, where we chose the latter.
As the training progresses, $\mathbf{a}_{S}$ learns to capture the importance of each context representation $\mathbf{c'}_i$.
For Att-RNN, we directly use its attention weights as $\mathbf{a}_S$.
For MHAtt-RNN, which adopts a multi-head attention mechanism, we are motivated by~\cite{vaswani2017attention} to choose one of the heads to yield attention weights.
While this approach is not universally applicable for arbitrary architectures, we emphasize that the only architectural requirement for the student model to satisfy is to output an intermediate features which preserve the temporal information.
There are already many prominent architectures that satisfy this requirement, e.g. CNN feature map before the global average pooling layer or sequence of outputs of RNN.

\subsection{Temporal Knowledge Distillation}\label{subsec:3_zero_overhead_attention_distillation}
To extract the temporal knowledge from the teacher model, we leverage $m$ self-attention maps $\mathbf{A}_{T}^{i=1,\cdots,m} \in \mathbb{R}^{n\times n}$ from $m$ transformer layers of the teacher model $\mathcal{F}_T$.
The attention rollout technique~\cite{DBLP:conf/acl/AbnarZ20} is applied to the self-attention maps to result in a single unified attention map $\mathbf{A}'_T \in \mathbb{R}^{n\times n}$.
We utilize the first vector $\mathbf{a}'_1\in\mathbb{R}^{1\times n}$ in $\mathbf{A}'_T$ as attention weights of the teacher model since the teacher is trained by performing the audio classification task based on the context representation $\mathbf{c}_1$ of the first time step.

To transfer the temporal knowledge from the teacher model to the student model, we align $\mathbf{a}'_1$ and $\mathbf{a}_S$ using the $\mathcal{L}_{KL}$ loss.
We define $\mathcal{L}_{KL}$ as a Kullback-Leibler (KL) divergence between the two attention weights so that minimizing the loss will penalize the misalignment.
However, the loss term $\mathcal{L}_{KL}$ cannot necessarily be directly computed since the dimensions of the two weights $\mathbf{a}'_1$ and $\mathbf{a}_S$ might not match ($n$ $\neq$ $n'$).
Therefore, we employ a simple linear interpolation method to resize the attention
weights to match the dimension of $\mathbf{a}'_1$ with $\mathbf{a}_S$ while preserving the temporal knowledge.
After applying the linear interpolation on the attention weights $\mathbf{a}'_1$, we obtain the final attention weights $\mathbf{a}_T$ of the teacher model $\mathcal{F}_{T}$.
Using the attention weights of the teacher and student models, the loss term $\mathcal{L}_{KL}$ is computed as follows:
\begin{equation}\label{eq:1_final_loss_term}
\mathcal{L}_{KL} = D_{KL}(\mathbf{a}_S|\mathbf{a}_T),
\end{equation}
where $D_{KL}$ is the KL divergence.

The final student loss $\mathcal{L}$ is defined as follows:
\begin{equation}\label{eq:1_final_loss_term}
\mathcal{L} = \lambda\mathcal{L}_{KL} + (1-\lambda)\mathcal{L}_{CLS},
\end{equation}
where the $\mathcal{L}_{CLS}$ is a cross-entropy-based classification loss of the student model.
$\lambda$ is a hyperparameter that controls the influence of each loss term.

Note that the on-device models (student models) except Att-RNN and MHAtt-RNN leverage the attention weights only during training to receive the knowledge from the teacher model, and they do not use the attention weights during inference.
In other words, there is no architectural change in the model during inference, hence showing zero computational overhead for the inference of on-device models.

\section{Experiments}\label{sec:3_experiment}
\subsection{Experimental Setup}\label{subsec:3_1_experimental_setup}
\textbf{Datasets.}
We verify the effectiveness of our proposed KD method with experiments on a real-world AED dataset (FSD50K) and a noisy KWS dataset (called noisy speech commands v2).

\textbf{\textit{FSD50K}}~\cite{fonseca2020fsd50k}:
The FSD50K dataset~\cite{fonseca2020fsd50k} is a multi-label audio event detection dataset, which represent real-world audios.
The dataset is composed of 51,197 human-labeled audio events with 200 classes with lengths ranging from 0.3 to 30 seconds. 
The audio inputs are zero-padded when the inputs are shorter than 30 seconds.

\textbf{\textit{Noisy Speech Commands v2}}:
To clearly demonstrate the effectiveness of our method in distilling the high-level knowledge of temporal information, we construct a noisy KWS dataset by inserting the existing KWS dataset, \textit{Speech Commands v2}~\cite{speechcommandsv2}, to the background speech noise.
The Speech Commands v2 dataset \cite{speechcommandsv2} contains 105,829 one-second utterances of 35 words stored as 16-bit mono PCM WAVE files with a 16KHz sample rate.
Following the settings from \cite{Rybakov_2020,speechcommandsv2}, we use their training splits and the 12 class labels, which include the ``\textit{silence}" label with no speech and ``\textit{unknown}" label with an additional 20 keywords.
We generate the synthetic audios by injecting the one-second speech command audios into the background speech noise obtained from the \textit{``Hubbub", ``speech noise", ``speech babble"} classes of the \textit{AudioSet} dataset~\cite{audioset17}.
The speech noise is randomly cropped to a predefined duration.
The locations of the speech command audios are uniformly sampled within the speech noise audio.
We define four Noisy Speech Commands datasets, each using a fixed duration of 2, 4, 6, and 8 seconds noise.
The noisy datasets are split using the same training and validation splits as the Speech Commands v2 dataset.
Audio mixing is done by weighted sum of both the one-second Speech Commands v2 and AudioSet noise PCM signals, with respective weights of $0.75$ and $0.25$.
\newline
\noindent
\textbf{Baselines.}
We employ the five on-device models described in Sec.~\ref{subsec:2_2_student_model} as baseline models.
By varying the hyperparameter $\lambda \in \{0.0, 0.1, 0.25, 0.5\}$, we observe the change of predictive performance to show the effect of the hyperparameter that controls the influence of our proposed method.
When $\lambda\text{=}0.0$, the loss becomes equal to the vanilla cross-entropy loss without knowledge distillation.
\newline
\noindent
\textbf{Metrics.}
Following~\cite{fonseca2020fsd50k} and~\cite{Rybakov_2020}, we use mean average precision (\textit{\textbf{mAP}}) and \textit{\textbf{accuracy}} to evaluate the performance on FSD50K and Noisy speech commands v2, which are the multi- and single-label classification datasets, respectively.
The higher score in both metrics indicates higher performance.
\newline
\noindent
\textbf{Implementation Details.}
The teacher model (XLSR-wav2vec 2.0) is pretrained on a multilingual speech dataset~\cite{conneau2020unsupervised}.
We fine-tune the teacher model with a batch size of 16 for 50 epochs.
For FSD50K and Noisy speech commands v2, we set the learning rate as 2e-5 and 5e-4, respectively, with 1K warmup steps.
We apply SpecAugment~\cite{DBLP:conf/interspeech/ParkCZCZCL19} with probability 0.75 which consists of two 10\% frequency mask while training.

The on-device models are trained for 20K iterations with a batch size of 100. 
% To optimize the baseline models, we use the SGD optimizer with a fixed learning rate of 1e-1 and momentum of 0.9.
Audios are re-sampled to a sample rate of 16kHz.
The student models take the MFCC representations of the audio as an input.
We employed a best keeping strategy to keep the weights with the best performance on the validation set using the evaluation metrics.
We leverage an existing code~\cite{Rybakov_2020} for our experiments.

\subsection{Evaluation Results}\label{subsec:3_2_evaluation_results}
\subsubsection{Results on Real-word AED dataset}\label{subsubsec:3_2_1_real_world_dataset}
\begin{table}[t]
\caption{
Performance comparison on the FSD50K dataset.
Test mAP of the best model found by the validation is reported, where the validation mAP is obtained every 400 steps.
}
\vspace{0.5em}
\label{tab:1_evaluation_results_fsd50k}
\scriptsize
\centering
% \resizebox{\linewidth}{!}{%
\begin{tabular}{c|c|ccc}
\toprule
\multirow{2}{*}{Model} & Vanilla & \multicolumn{3}{c}{Attention Distillation} \\
 & $\lambda=0.0$ & $\lambda=0.1$ & $\lambda=0.25$ & $\lambda=0.5$ \\
\midrule

wav2vec 2.0 & 0.5498 & \multicolumn{3}{c}{ N/A } \\
\cmidrule{1-5}
LSTM-P & 0.1141 & 0.1274 & \textbf{ 0.1300 } & 0.1043 \\
TC-ResNet & 0.1814 & 0.1841 & \textbf{ 0.1951 } & 0.1509 \\
CRNN & 0.2789 & 0.2670 & 0.2835 & \textbf{ 0.3053 } \\
Att-RNN & 0.2856 & \textbf{ 0.3471 } & 0.2885 & 0.2891 \\
MHAtt-RNN & 0.2647 & 0.1694 & 0.3182 & \textbf{ 0.3317 } \\
\bottomrule
\end{tabular}
\vspace{-2em}
\end{table}

Table~\ref{tab:1_evaluation_results_fsd50k} demonstrates the effectiveness of our method on FSD50K dataset.
Our method shows clear improvement on all the representative architectures we have chosen, where the mAP scores of LSTM-P, TC-ResNet, CRNN, Att-RNN, and MHAtt-RNN increased by 13.9\%, 7.6\%, 9.5\%, 21.5\%, and 25.3\%, respectively.
Especially, improvement on attention-based methods is substantial.

\subsubsection{Results on Noisy KWS dataset}\label{subsubsec:3_2_2_synthetic_dataset}
\begin{table}[t]
\caption{
Performance comparison on Noisy Speech Commands v2 dataset.
Test accuracy (\%) of the best model found by the validation accuracy is reported, where the validation accuracy is obtained every 400 steps.
Best accuracies are in bold, and the performance of the student models that outperform the teacher model is underlined.
}
\vspace*{0.3cm}
\label{tab:2_evaluation_results_noisy}

\scriptsize
\centering
% \resizebox{0.47\textwidth}{!}{%

\begin{tabular}{c|c|c|ccc}

\toprule

Audio & \multirow{2}{*}{Model} & Vanilla & \multicolumn{3}{c}{Attention Distillation} \\
Length &  & $\lambda=0.0$ & $\lambda=0.1$ & $\lambda=0.25$ & $\lambda=0.5$ \\
\midrule

\multirow{6}{*}{ 2s } & wav2vec 2.0 & 90.59 & \multicolumn{3}{c}{ N/A } \\
\cmidrule{2-6}
& LSTM-P & 88.73 & 88.98 & \textbf{ 89.31 } & 88.92 \\
& TC-ResNet & 87.77 & \textbf{ 88.08 } & 86.21 & 86.27 \\
& CRNN & 89.96 & \textbf{ 90.06 } & 90.00 & 89.46 \\
& Att-RNN & 89.88 & \underline{91.31} & \textbf{ \underline{91.67} } & \underline{90.94} \\
& MHAtt-RNN & 89.75 & \underline{91.25} & \underline{91.67} & \textbf{ \underline{91.75} } \\
\midrule

\multirow{6}{*}{ 4s } & wav2vec 2.0 & 91.22 & \multicolumn{3}{c}{ N/A } \\
\cmidrule{2-6}
& LSTM-P & 85.19 & 88.23 & 88.52 & \textbf{ 89.08 } \\
& TC-ResNet & 87.60 & \textbf{ 88.33 } & 87.21 & 84.62 \\
& CRNN & 89.69 & 89.44 & 89.46 & \textbf{ 90.21 } \\
& Att-RNN & 90.65 & \textbf{ \underline{91.98} } & \underline{91.79} & \underline{91.58} \\
& MHAtt-RNN & 91.19 & \underline{91.58} & \textbf{ \underline{92.12} } & \underline{91.73} \\
\midrule

\multirow{6}{*}{ 6s } & wav2vec 2.0 & 90.93 & \multicolumn{3}{c}{ N/A } \\
\cmidrule{2-6}
& LSTM-P & 45.27 & 69.21 & \textbf{ 85.58 } & 85.10 \\
& TC-ResNet & 86.00 & \textbf{ 86.85 } & 84.23 & 82.10 \\
& CRNN & 88.58 & \textbf{ 89.88 } & 89.29 & 89.46 \\
& Att-RNN & 90.88 & 90.77 & 90.73 & \textbf{ \underline{91.19} } \\
& MHAtt-RNN & 90.58 & \underline{90.96} & \textbf{ \underline{91.67} } & \underline{91.10} \\
\midrule

\multirow{6}{*}{ 8s } & wav2vec 2.0 & 90.95 & \multicolumn{3}{c}{ N/A } \\
\cmidrule{2-6}
& LSTM-P & 78.44 & \textbf{ 82.19 } & 34.58 & 66.25 \\
& TC-ResNet & 77.81 & \textbf{ 85.79 } & 85.71 & 80.15 \\
& CRNN & 88.94 & 89.02 & \textbf{ 89.79 } & 89.77 \\
& Att-RNN & 88.81 & 90.44 & \textbf{ \underline{90.98} } & 90.75 \\
& MHAtt-RNN & 88.33 & \underline{91.50} & \textbf{ \underline{91.79} } & \underline{91.35} \\
\bottomrule

\end{tabular}
\vspace{-1em}

% }
\end{table}

Table~\ref{tab:2_evaluation_results_noisy} summarizes the evaluation results on the Noisy Speech Commands v2.
Our method shows superior performance across all datasets on all the student architectures.
Furthermore, attention-based student models that applied attention distillation often exceed the teacher model performance, whereas vanilla student models are always inferior to the teacher model.
We also observed that the accuracy disparity between models trained with vanilla and attention distillation losses tends to grow for non-attention-based models as the audio length increases.
This implies that those models benefit from the temporal knowledge extracted from attention weights without any architectural changes.

\subsubsection{Further Analysis}\label{subsubsec:3_2_3_further_analyses}
\begin{figure}[t] %%% t: top, b: bottom, h: here
\centering
\includegraphics[width=0.97\columnwidth]{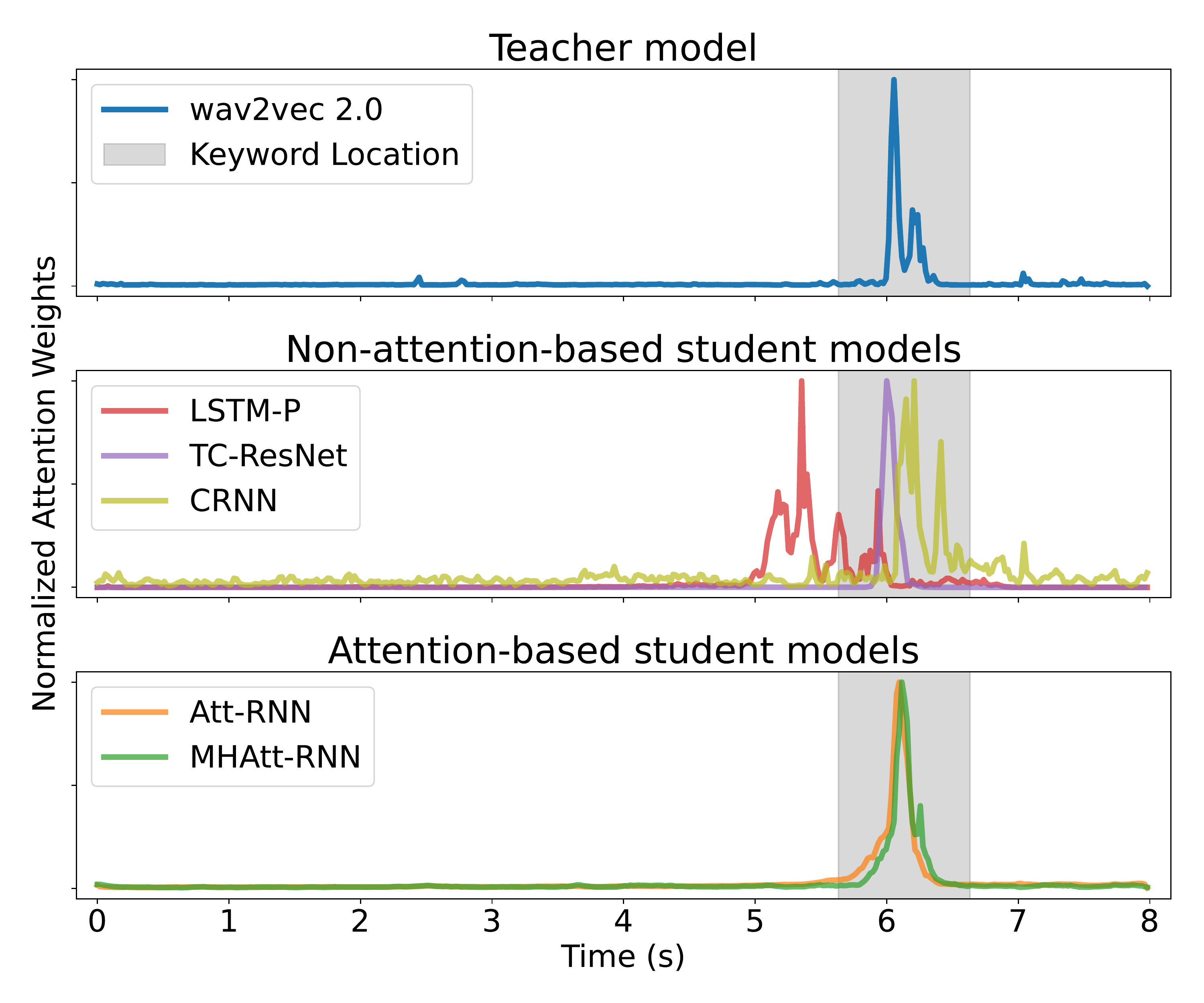}
\vspace{-1em}
\caption{
Visualization of attention weights extracted from multiple models.
We input an arbitrary sample from the Noisy Speech Commands v2 dataset with 8 seconds noise.
We plot the location of the one second keyword to all the plots.
}
\label{fig:2_attention_visualization}
\end{figure}
In Figure~\ref{fig:2_attention_visualization}, we visualize attentions extracted from the teacher model and all the student models applied on 8 second audio sample from Noisy Speech Commands v2.
Student architectures are selected based on the best categorical accuracy from Table~\ref{tab:2_evaluation_results_noisy}.
The gray region represents the position of the keyword location within the audio.
We observe that even though the teacher model is trained only with the classification label, attention weights successfully focuses on the keyword location.
We can also see that all the on-device models attend at similar positions inside the keyword location, indicating that the teacher and the student attention weights are accurately aligned.
\section{Conclusion}\label{sec:4_conclusion}
In this paper, we propose a novel attention distillation method that transfers the temporal knowledge from large teacher models to on-device student audio classification models.
We extract the attention weights from both the teacher and the student models, and align them via KL divergence.
Our method can be applied to various architectures with no architectural change during inference.
Through extensive experiments on both an audio event detection dataset and a noisy keyword spotting dataset, we show that our proposed method improves the predictive performance.
Through extensive experiments on both an audio event detection dataset and a noisy keyword spotting dataset, we show that our proposed method improves the predictive performance.

\bibliographystyle{IEEEbib}
\bibliography{references}

\begin{thebibliography}{10}

\bibitem{speechcommandsv2}
P.~{Warden},
\newblock ``{Speech Commands: A Dataset for Limited-Vocabulary Speech
  Recognition},''
\newblock {\em Proc. IEEE ICASSP}, Apr. 2018.

\bibitem{Rybakov_2020}
Oleg Rybakov, Natasha Kononenko, Niranjan Subrahmanya, Mirkó Visontai, and
  Stella Laurenzo,
\newblock ``Streaming keyword spotting on mobile devices,''
\newblock {\em INTERSPEECH}, Oct 2020.

\bibitem{fonseca2020fsd50k}
Eduardo Fonseca, Xavier Favory, Jordi Pons, Frederic Font, and Xavier Serra,
\newblock ``{FSD50K}: an open dataset of human-labeled sound events,''
\newblock {\em arXiv preprint arXiv:2010.00475}, 2020.

\bibitem{FutamiIUMSK20}
Hayato Futami, Hirofumi Inaguma, Sei Ueno, Masato Mimura, Shinsuke Sakai, and
  Tatsuya Kawahara,
\newblock ``Distilling the knowledge of {BERT} for sequence-to-sequence
  {ASR},''
\newblock in {\em INTERSPEECH}, Oct 2020.

\bibitem{Lu2017KnowledgeDF}
Liang Lu, Michelle Guo, and Steve Renals,
\newblock ``Knowledge distillation for small-footprint highway networks,''
\newblock {\em Proc. IEEE ICASSP}, pp. 4820--4824, 2017.

\bibitem{hinton2015distilling}
Geoffrey Hinton, Oriol Vinyals, and Jeff Dean,
\newblock ``Distilling the knowledge in a neural network,''
\newblock {\em arXiv preprint arXiv:1503.02531}, 2015.

\bibitem{berg21_interspeech}
Axel Berg, Mark O'Connor, and Miguel~Tairum Cruz,
\newblock ``{Keyword Transformer: A Self-Attention Model for Keyword
  Spotting},''
\newblock in {\em INTERSPEECH}, 2021, pp. 4249--4253.

\bibitem{MunimIS19}
Raden~Mu’az Mun’im, Nakamasa Inoue, and Koichi Shinoda,
\newblock ``Sequence-level knowledge distillation for model compression of
  attention-based sequence-to-sequence speech recognition,''
\newblock in {\em Proc. IEEE ICASSP}. IEEE, 2019, pp. 6151--6155.

\bibitem{codert21}
Rupak~Vignesh Swaminathan, Brian King, Grant~P. Strimel, Jasha Droppo, and
  Athanasios Mouchtaris,
\newblock ``Codert: Distilling encoder representations with co-learning for
  transducer-based speech recognition,''
\newblock {\em arXiv preprint arXiv:2106.07734}, 2021.

\bibitem{vaswani2017attention}
Ashish Vaswani, Noam Shazeer, Niki Parmar, Jakob Uszkoreit, Llion Jones,
  Aidan~N Gomez, {\L}ukasz Kaiser, and Illia Polosukhin,
\newblock ``Attention is all you need,''
\newblock in {\em Advances in neural information processing systems}, 2017, pp.
  5998--6008.

\bibitem{Zagoruyko2017pay}
Sergey Zagoruyko and Nikos Komodakis,
\newblock ``Paying more attention to attention: Improving the performance of
  convolutional neural networks via attention transfer,''
\newblock {\em ICLR}, 2017.

\bibitem{Chang2020IntraUtteranceSP}
Chun-Chieh Chang, Chieh-Chi Kao, Ming Sun, and Chao Wang,
\newblock ``Intra-utterance similarity preserving knowledge distillation for
  audio tagging,''
\newblock in {\em INTERSPEECH}, 2020.

\bibitem{conneau2020unsupervised}
Alexis Conneau, Alexei Baevski, Ronan Collobert, Abdelrahman Mohamed, and
  Michael Auli,
\newblock ``Unsupervised cross-lingual representation learning for speech
  recognition,''
\newblock {\em arXiv preprint arXiv:2006.13979}, 2020.

\bibitem{audioset17}
Jort~F. Gemmeke, Daniel P.~W. Ellis, Dylan Freedman, Aren Jansen, Wade
  Lawrence, R.~Channing Moore, Manoj Plakal, and Marvin Ritter,
\newblock ``Audio set: An ontology and human-labeled dataset for audio
  events,''
\newblock in {\em Proc. IEEE ICASSP}, 2017.

\bibitem{devlin2019bert}
Jacob Devlin, Ming-Wei Chang, Kenton Lee, and Kristina Toutanova,
\newblock ``Bert: Pre-training of deep bidirectional transformers for language
  understanding,''
\newblock in {\em NAACL-HLT}, 2019.

\bibitem{DBLP:journals/jmlr/GersSS02}
Felix~A. Gers, Nicol~N. Schraudolph, and J{\"{u}}rgen Schmidhuber,
\newblock ``Learning precise timing with {LSTM} recurrent networks,''
\newblock {\em J. Mach. Learn. Res.}, vol. 3, pp. 115--143, 2002.

\bibitem{Choi2019tcresnet}
Seungwoo Choi, Seokjun Seo, Beomjun Shin, Hyeongmin Byun, Martin Kersner,
  Beomsu Kim, Dongyoung Kim, and Sungjoo Ha,
\newblock ``Temporal convolution for real-time keyword spotting on mobile
  devices,''
\newblock in {\em INTERSPEECH}, Sep 2019.

\bibitem{DBLP:conf/interspeech/ArikKCHGFPC17}
Sercan~{\"{O}}mer Arik, Markus Kliegl, Rewon Child, Joel Hestness, Andrew
  Gibiansky, Christopher Fougner, Ryan Prenger, and Adam Coates,
\newblock ``Convolutional recurrent neural networks for small-footprint keyword
  spotting,''
\newblock in {\em INTERSPEECH}, Aug 2017.

\bibitem{Andrade2018aneural}
D.~{Coimbra de Andrade}, S.~{Leo}, M.~{Loesener Da Silva Viana}, and
  C.~{Bernkopf},
\newblock ``{A neural attention model for speech command recognition},''
\newblock {\em arXiv preprint arXiv:1808.08929}, 2018.

\bibitem{yang2016hierarchical}
Zichao Yang, Diyi Yang, Chris Dyer, Xiaodong He, Alex Smola, and Eduard Hovy,
\newblock ``Hierarchical attention networks for document classification,''
\newblock in {\em NAACL-HLT}, 2016, pp. 1480--1489.

\bibitem{DBLP:conf/acl/AbnarZ20}
Samira Abnar and Willem~H. Zuidema,
\newblock ``Quantifying attention flow in transformers,''
\newblock in {\em ACL}, 2020.

\bibitem{DBLP:conf/interspeech/ParkCZCZCL19}
Daniel~S. Park, William Chan, Yu~Zhang, Chung{-}Cheng Chiu, Barret Zoph,
  Ekin~D. Cubuk, and Quoc~V. Le,
\newblock ``Specaugment: {A} simple data augmentation method for automatic
  speech recognition,''
\newblock in {\em INTERSPEECH}, 2019.

\end{thebibliography}

% \appendix
% \input{Sections/A_Rebuttal}

\end{document}